\documentclass[aps,pra,twocolumn,floatfix,showpacs]{revtex4-1}
\usepackage{graphicx,amsfonts}
\usepackage{amsmath,amscd,amsfonts,amssymb,color}
\newcommand{\bra}[1]{\left\langle{#1}\right\vert}
\newcommand{\ket}[1]{\left\vert{#1}\right\rangle}

\begin{document}

\title{Can Two Quantum Cheshire Cats Exchange Grins?}

\author{Debmalya Das\(^{1,2,3}\) and Arun Kumar Pati\(^{3}\)}

\affiliation{\(^1\)Department of Physics and Astronomy, University of Rochester, Rochester, New York 14627, USA\\
\(^2\)Center for Coherence and Quantum Optics, University of Rochester, Rochester, New York 14627, USA.\\
\(^3\)Quantum Information and Computation Group, Harish-Chandra Research Institute, HBNI, Chhatnag Road, Jhunsi,
Allahabad 211 019, India.\\}


\begin{abstract}
A common-sense perception of a physical system is that it is inseparable from its physical properties.
The notion of Quantum Cheshire Cat challenges this, as far as quantum systems are concerned. It shows that 
a quantum system can be decoupled from its physical property under suitable pre and postselections. 
However, in the Quantum Cheshire Cat setup, the decoupling is not permanent. The photon, for example,
and its circular polarization is separated and then recombined. In this paper, we present a thought experiment
where we decouple two photons from their respective polarizations and then interchange them during
recombination. Thus, our proposal shows that that the belongingness of a property for a physical system is
very volatile in the quantum world. This raises the question of reality of an observable at a much deeper
level.
\end{abstract}

\maketitle


{\it Introduction.-}
It is commonly believed that the properties of a physical system cannot be separated from the system
itself. This picture of the nature of physical systems, however, does not hold true in the realm of
quantum mechanics. A thought experiment, known as the Quantum Cheshire Cat~\cite{Cheshire}, shows that a
property such as the polarization of a photon can exist in isolation to the photon itself. Based on 
a modified version of a Mach-Zehnder interferometer, the Quantum Cheshire Cat demonstrates that a 
photon and its circular polarization can be decoupled from each other and made to travel separately 
through the two arms. This echoes the description of certain events in the novel Alice in Wonderland
where Alice remarks,``Well! I've often seen a cat without a grin,..., but a grin without a cat!
It's the most curious thing I ever saw in my life!''~\cite{Carroll1865}.

Quantum Cheshire Cat has opened up a new window for understanding of quantum systems~\cite{Bancal2013,
Duprey2018, Correa2015, Atherton2015}. 
It pertains not only to photons
and their polarizations but can, in principle, be observed with any quantum system and its property,
such as neutron and its magnetic moment, electron and its charge and so on. Experimental verifications 
of the phenomenon with neutron as the cat and its magnetic moment as the grin have been conducted
~\cite{Denkmayr2014, Sponar2016}. The phenomenon has also been observed experimentally in the context of
photon and polarization~\cite{Ashby2016}. Further developments on the idea of the Quantum Cheshire Cat include
the proposal of a complete Quantum Cheshire Cat~\cite{CC} and twin Quantum Cheshire Cats~\cite{Twin}.
The effect has been used to realize the three box paradox~\cite{Pan2013} and has been studied in the
presence of decoherence~\cite{Richter2018}. Recently, a protocol has been developed using which the decoupled
grin of a Quantum Cheshire Cat has been teleported between two spatially separated parties without the cat
~\cite{Das2019}.

There also exist criticisms of Quantum Cheshire Cat in literature. A comprehensive list of the 
same can be found in Ref.~\cite{Duprey2018}. Typically, the critics 
tend to refute the claim of the disembodiment of the polarization from the photon, using different types of 
arguments. In Ref.~\cite{Stuckey2015}, the absence of the polarization in a particular arm, where the 
corresponding weak value is zero, is dismissed due to a particular form of the intensity obtained at the detector.
However, in the analysis, the authors do not consider weak measurements at all, which we discussed, is vital
for the observation of Quantum Cheshire Cat. In another work~\cite{Atherton2015}, the results of the neutron interferometry 
experiment~\cite{Denkmayr2014}, demonstrating Quantum Cheshire Cat, is mimicked by using classical beams and observing electric
field intensities at the output detectors. The authors strive
to show that there is nothing particularly "quantum" in the Quantum Cheshire Cat. Yet, the reasoning suffers from the 
same shortcoming of the previous example. The logic of the phenomenon is heavily dependent on defining weak values
of quantum observables like spin or polarization. In Ref.~\cite{Atherton2015}, a precise definition of such quantum observables
is found wanting, as one would expect would be the case for classical beams. Another argument is that Quantum 
Cheshire Cat is just a form of quantum interference of a particular kind of meter states. The problem of this 
approach is that it talks about the meter state dynamics and does not involve an interaction Hamiltonian,
resulting in no weak values~\cite{Correa2015}. Also the numerical simulations in Ref.~\cite{Michielsen2015}, the target 
of criticism is the
neutron interferometry experiment, meant to be experimentally demonstrating the effect, and therefore in no way
the negative results actually disapprove the disembodiment in the original theoretical proposal.

The premise of the realization of the Quantum Cheshire Cat involves weak measurements and weak values~\cite{AAV1988}.
The development of the concept weak measurements and weak values stemmed from the limitation 
posed by the measurement problem in quantum mechanics, in acquiring knowledge about the value 
of an observable of a quantum system. Given a quantum system in a general pure state $\ket{\Psi}$, if a 
measurement of an observable $A$ is performed, then the outcome can be any one of the eigenvalues $a_n$ of 
$A$. Measurement of $A$ on an ensemble of states, shows that the measurement outcomes are indeterministic
and probabilistic, the probability of an outcome $a_n$ in any given run being 
$|\langle a_n\ket{\Psi}|^2$, where $\ket{a_n}$ is the eigenstate,
corresponding to the eigenvalue $a_n$.
After the strong measurement, the state of the system collapses to 
$\ket{a_n}$. As a result, there is no general consensus among physicists as to whether the measurement
of an observable actually reveals a property of the system or is an artifact of the measurement process
itself. The proponents of weak value tried to circumvent this problem of wavefunction collapse by 
minimizing the disturbance caused to the initial state. This is achieved by weakly measuring the 
observable $A$, causing minimal disturbance to the state. We briefly recapitulate the process of weak
measurement below.

Consider a quantum system preselected in the state $\ket{\Psi_i}$. Suppose an observable $A$ is weakly measured
by introducing a small coupling between the quantum system and a suitable measurement device or a meter.
A second observable $B$ is thereafter measured strongly and one of its eigenstates $\ket{\Psi_f}$ is postselected.
For all successful postselections of the
state $\ket{\Psi_f}$, the meter readings corresponding to the weak
measurements of $A$ are taken into consideration while the
others are discarded. The shift in the meter readings, on
an average, for all such postselected systems is given by the weak value $A_w$~\cite{AAV1988}.
Mathematically the weak value of $A$ is defined as
\begin{equation}
 A_w=\frac{\bra{\Psi_f}A\ket{\Psi_i}}{\bra{\Psi_f}\Psi_i\rangle}.
 \label{wv_def}
\end{equation}

The weak value is therefore to be interpreted as the value the observable A, takes
between the preselected state $\ket{\Psi_i}$ and the postselected state $\ket{\Psi_f}$.
The measurement of $A$ must be weak to preserve the initial probability distribution for the 
final postselected state. Weak values via weak measurements have been observed experimentally~\cite{Pryde2005}.
The weak value of an observable can be complex~\cite{Jozsa2007} or can take up large values
that lie outside the eigenvalue spectrum~\cite{AAV1988, Duck1989, Tollaksen2010}. The latter has led to the use of weak 
measurements as a tool for signal amplification~\cite{Dixon2009, Nishizawa2012}. A geometrical interpretation of
weak values can be found in Ref.~\cite{Cormann2016}. Among the myriad applications of 
weak measurements are observation of the spin Hall effect~\cite{Hosten2008}, resolving quantum paradoxes~\cite{Hofmann2010,
Dolev2005}, quantum state visualization~\cite{Kobayashi2014}, quantum state tomography~\cite{Hofmann2010, Wu2013},
direct measurement of wavefunction~\cite{Lundeen2011, Lundeen2012}, probing contextuality~\cite{Tollaksen2007}
, measuring the expectation value of non-Hermitian operators~\cite{Pati2015, Nirala2019} and quantum precision thermometry
~\cite{Pati2019}.

Any probing of position or a component of polarization of the photon for the observation of the Quantum Cheshire Cat 
effect must necessarily be weak. This is because projective measurements tend to destroy the original state 
of a system while extracting information. Weak measurements, on the other hand, minimally disturb the system
while gaining small information about it. The state cannot be disturbed in a Quantum Cheshire Cat setup as any 
alteration of the original state will lead to altered probabilities of the postselected state.

In this paper we explore yet another counterintuitive aspect of the Quantum Cheshire Cat. We design a setup
where we not only decouple the grin (circular polarization) from the Quantum Cheshire Cat (photon), but also 
replace it with a grin (circular polarization)
originally belonging to another Cheshire Cat (a different photon). Our setup is comprised of two modified and overlapping 
Mach-Zehnder interferometers. We try to show that a property of a physical system does
not uniquely belong to that system in the quantum domain and can be replaced by the same property from another
physical system. Following Neils Bohr, the reality of a physical attribute comes into existence only after a
measurement. Our result shows that physical attributes are not real and may not belong to a system under 
certain scenario. It is to be noted that this exchange of polarizations of two photons is not merely an exchange of 
two states of polarizations as in spin swapping. Here we literally strip the photon of its polarization and then exchange
it with another polarization. Thus, our result raises the question of reality of observables at a much deeper level 
than what Bohr's famous quote suggests. 



{\it The Quantum Cheshire Cat.-}
In 2013, Aharonov et. al.~\cite{Cheshire} proposed a thought experiment where it is possible to decouple a
property of a system from the system itself. The thought experiment can be realized by a scheme that is based
on a Mach-Zehnder interferometer and pertains to the separation of a photon from its circular polarization.
To begin with, a source sends a linearly polarized single photon towards a 50:50 beam-splitter 
$BS_1$. Let us denote the left and the right paths of the photon, after passing the beam-splitter, as 
$\ket{L}$ and $\ket{R}$, respectively which are the two orthogonal states of the photon. If the photon, 
from the source, 
is initially in a horizontal polarization state $\ket{H}$, after passing through the beam-splitter $BS_1$ it
goes to the state 
\begin{equation}
 \ket{\Psi_i}= \frac{1}{\sqrt{2}}(i\ket{L}+\ket{R})\ket{H}),
\end{equation}
where the relative phase factor $i$ appears as the photon traveling through the left arm  is reflected by the 
beam splitter. In the analysis of the Quantum Cheshire Cat, this state is used as the preselected state. A 
combination of a half-waveplate (HWP), a phase-shifter (PS), beam-splitter $BS_2$, a 
polarization beam-splitter (PBS) and three detectors $D_1$, $D_2$ and $D_3$ conducts the postselection. Let us 
call this the postselection block. The required postselected state to observe the Quantum Cheshire Cat phenomenon 
is given by
\begin{equation}
 \ket{\Psi_f}=\frac{1}{\sqrt{2}}(\ket{L}\ket{H}+\ket{R}\ket{V}),
\end{equation}
where $\ket{V}$ refers to the vertical polarization state orthogonal to the initial polarization state 
$\ket{H}$.
To appreciate how the postselection works, note that the HWP flips the polarization of the photon from
$\ket{H}$ to $\ket{V}$ and vice-versa. The phase-shifter (PS) adds a phase factor of $i$
to the state of the photon passing through it. The beam-splitter $BS_2$ allows only a photon
in the superposition state $\frac{1}{\sqrt{2}}(\ket{L}+i\ket{R})$ towards the PBS and never towards $D_2$.
Next, the PBS always transmits the horizontal polarization
$\ket{H}$ and always reflects the vertical polarization $\ket{V}$. Thus only if a state $\ket{\Psi_f}$ 
enters the postselection block, there is a click of detector $D_1$. Any clicking of the detectors $D_2$ or
$D_3$ implies a different state entering the postselection block. Therefore, by selecting the clicks of the 
detector $D_1$ alone and discarding all the others, one can postselect the state $\ket{\Psi_f}$.

\label{recap}
\begin{figure}
 \includegraphics[scale=0.8]{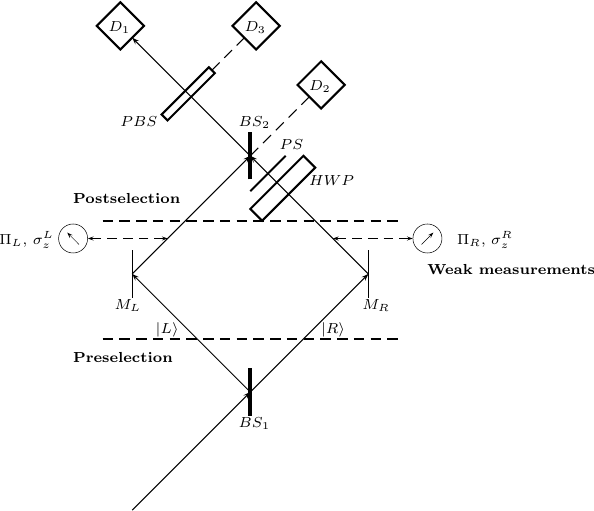}
 \caption{The basic Quantum Cheshire Cat setup. The initial state $\ket{\Psi}$ is prepared by passing a 
 photon with linear polarization $\ket{H}$ through a beam-splitter $BS_1$. Weak measurements of positions
 and photon polarizations are carried out on the two arms of the interferometer. The postselection block 
 consists of a half-waveplate $HWP$, a phase-shifter $PS$, the beam-splitter $BS_2$, a polarization beam-
 splitter $PBS$ that transmits polarization states $\ket{H}$ and reflects state $\ket{V}$ and three detectors
 $D_1$, $D_2$ and $D_3$.}
\end{figure}

%
%

To find out which arm a photon passed through, we must perform measurements of the observables 
$\Pi_L=\ket{L}\bra{L}$ and $\Pi_R=\ket{R}\bra{R}$ in the left and the right arms, respectively. However, we 
must also ensure that the measurement does not disturb the initial state much, so that the probability statistics
for the desired postselected state is minimally altered. To this end the observables $\Pi_L=\ket{L}\bra{L}$ and 
$\Pi_R=\ket{R}\bra{R}$ must be weakly measured by placing weak
detectors in the two arms. Similarly, the polarizations can be detected in the left and the right arms
by respectively performing weak measurements of the following operators
\begin{eqnarray}
 \sigma^L_z=\Pi_L \otimes\sigma_z,\nonumber\\
 \sigma^R_z=\Pi_R \otimes\sigma_z,
\end{eqnarray}
where
\begin{equation}
 \sigma_z=\ket{+}\bra{+}-\ket{-}\bra{-},
\end{equation}
the circular polarization basis $\{\ket{+},\ket{-}\}$ itself defined as
\begin{eqnarray}
 \ket{+}&=&\frac{1}{\sqrt{2}}(\ket{H}+ i \ket{V}),\nonumber\\
 \ket{-}&=&\frac{1}{\sqrt{2}}(\ket{H}- i \ket{V}).
\end{eqnarray} 
The weak values of the photon positions are measured to be
\begin{equation}
 (\Pi_L)^w=1 \;\mbox{and}\; (\Pi_R)^w=0
 \label{position_wv}
\end{equation}
which implies that the photon has traveled through the left arm. The weak 
values of the polarization positions, on the other hand, are measured to be
\begin{equation}
 (\sigma^L_z)^w=0\;\mbox{and}\;(\sigma^R_z)^w=1.
 \label{polarization_wv}
\end{equation}
Equations~(\ref{position_wv}) and~(\ref{polarization_wv}) jointly tell us that the photon traveled 
through the left arm but its circular polarization traveled through the right arm. This means
the two degrees of freedom of a single entity have been decoupled. Thus, the polarization, a property
of the photon, can exist independent of the existence of the photon, in that region.

 {\it Exchanging Grins of Cats.--} 
 In the previous section we have seen that the grin of a cat can be separated from the cat itself
 under suitable choices of pre and postselection. Here we consider two such Quantum Cheshire Cats and
 exchange their grins. In terms of physical realization, we take two linearly polarized photons, decouple their circular
 polarizations and then recouple them with the other photon.
 
\begin{widetext}
 \begin{figure}
 \includegraphics[scale=1]{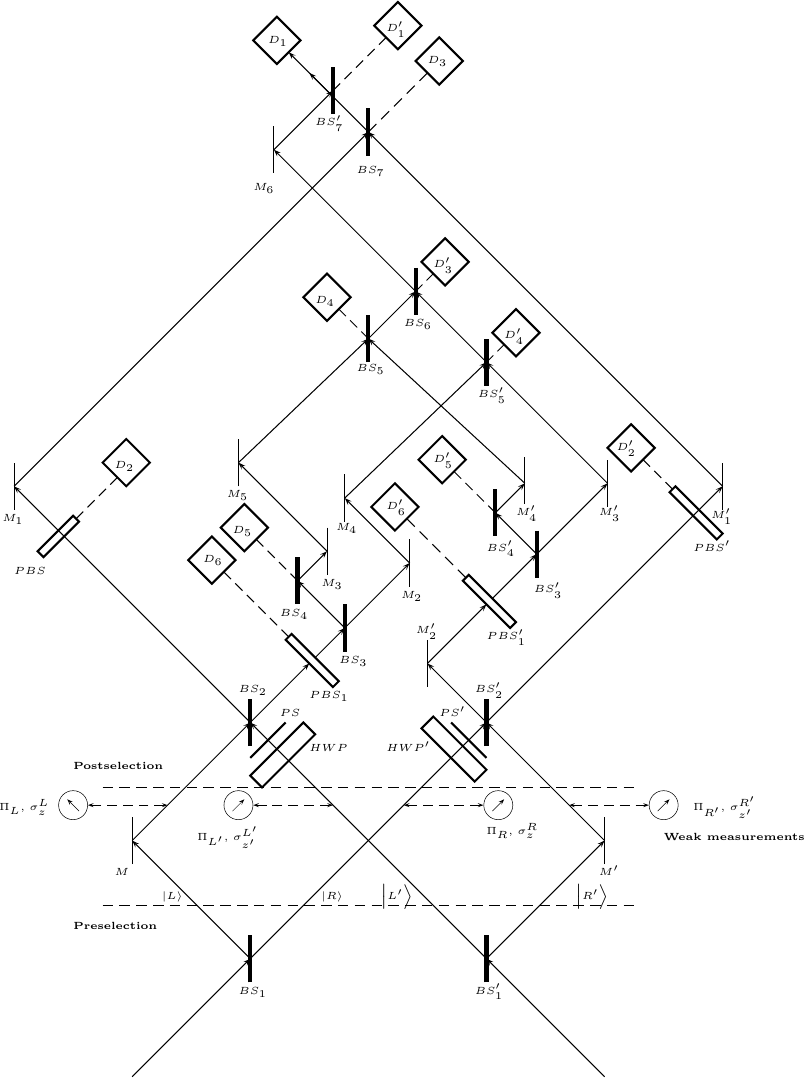}
 \caption{Exchange of the grins of two Quantum Cheshire Cats. The desired postselection for observing the 
 exchange of the grins is obtained by selecting only the cases for which there are clicks of the detector
 $D_1$. Several mirrors have been used just to accommodate the detectors in a compact space.\label{CC_Swap}}
\end{figure}
 \end{widetext}

 To see this effect let us consider the setup shown in Fig.~\ref{CC_Swap}. There are two sources of linearly 
 polarized photons,
 near the unprimed and the primed halves of the arrangement, on the left and the right, respectively. The input
 polarization in the unprimed half is $\ket{H}$ while that in the primed half is $\ket{H^\prime}$. The setup 
 allows one to prepare a preselected state given by

\begin{eqnarray}
 \ket{\Psi_i}&=&\frac{1}{2}(i\ket{L}\ket{H}\ket{L^\prime}\ket{H^\prime}+\ket{R}\ket{H}\ket{L^\prime}\ket{H^\prime}\nonumber\\
 &&-\ket{L}\ket{H}\ket{R^\prime}\ket{H^\prime}+i\ket{R}\ket{H}\ket{R^\prime}\ket{H^\prime}).
 \label{prs}
\end{eqnarray}

where $\ket{L}$ and $\ket{R}$ indicate the states of a photon in the left and right arms of the unprimed half
of the apparatus
and $\ket{L^\prime}$ and $\ket{R^\prime}$ indicate the states of a photon in the left and right arms of the 
primed half of the apparatus. The right arm of the unprimed half is connected to the output of the primed half
and the left arm of the primed half is connected to the output of the unprimed half.

Now, the choice of $\ket{\Psi_f}$ which is constrained by the need to define the weak values,
makes it necessary to chose a
rather  a complicated post-selection strategy. In the present work, we use a four qubit entangled state  for our purpose. 
This may look slightly complicated procedure and other choices of post-selection with simple realization may be possible,
but we leave that for future. 
The photons and their polarizations are postselected in the state as given by
\begin{eqnarray}
 \ket{\Psi_f}&=&\frac{1}{2}(\ket{L}\ket{H}\ket{R^\prime}\ket{H^\prime}+\ket{R}\ket{V}\ket{R^\prime}\ket{H^\prime}+\nonumber\\
 &&\ket{L}\ket{H}\ket{L^\prime}\ket{V^\prime}-\ket{R}\ket{V}\ket{L^\prime}\ket{V^\prime}).\label{pos}
\end{eqnarray}
This a four qubit maximally entangled state and is one of the cluster states~\cite{Briegel2001}. As we shall
see the choice of this state leads to interesting consequences which is the subject of this paper.
The postselection is realized using 
the following setup. Beam-splitters $BS_i (i= 2,3,4,5,6,7 )$ and $BS^{\prime}_i (i = 2,3,4,5,7) $,
half-waveplates $HWP$ and $HWP^\prime$, phase-shifters
$PS$ and $PS^\prime$, four polarization beam-splitters $PBS$, $PBS^\prime$, $PBS_1$ and $PBS^\prime_1$ and detectors
$D_i (i = 1,..,6)$ and $D^{\prime}_i (i = 1,..,6)$ together constitute the postselection block.
It is clear that this postselected state $\ket{\Psi_f}$ demands that there is entanglement
between the path degrees of freedom of the two halves of the interferometric arrangement. The $HWP$ 
and the $HWP^\prime$ cause the transformations $\ket{H^\prime}\leftrightarrow\ket{V^\prime}$ and
$\ket{H}\leftrightarrow\ket{V}$, respectively, and $PS$ and $PS^\prime$ add a phase-factor $i$.
Now let the beam-splitters $BS_2$ and $BS_2^\prime$ are chosen such that if a state 
$\ket{L}\ket{L^\prime}$ is incident on $BS_2$ or a state $\ket{R}\ket{R^\prime}$ is incident on $BS_2^\prime$,
then the photons emerge towards $PBS$ and $PBS^\prime$, respectively. The $PBS$ and the $PBS^\prime$, once again, 
allow only polarizations $\ket{H^\prime}$ and $\ket{H}$ to be transmitted and other polarizations to be reflected.
Now if any other state is incident on $BS_2$ and $BS^\prime_2$, it proceeds towards $PBS_1$ or $PBS^\prime_1$.
These polarization beam splitters, once again, allow polarizations $\ket{H^\prime}$ and $\ket{H}$ to be transmitted,
towards $BS_3$ and $BS^\prime_3$, respectively, and reflect polarizations  $\ket{V^\prime}$ and $\ket{V}$ towards
$D_6$ and $D^\prime_6$. Next, the beam-splitter $BS_3$ is so chosen that the state $\ket{L}$ is transmitted
towards the mirror $M_2$ then $M_4$ and finally towards the beam-splitter $BS^\prime_5$ while any other state is reflected 
towards $BS_4$. In conjunction with this, the 
beam-splitter $BS^\prime_3$ is chosen to transmit the state $\ket{R^\prime}$ towards the mirror $M^\prime_3$ and then towards
the beam-splitter $BS^\prime_5$ and reflect any other 
state towards $BS^\prime_4$. The beam-splitter $BS^\prime_5$ receives inputs $\ket{L}$ and $\ket{R^\prime}$. It must be 
chosen such that a state $\ket{L}\ket{R^\prime}$ is channeled towards the beam-splitter $BS_6$ while any other state is 
sent to the detector $D^\prime_4$. On the other hand, the beam-splitter $BS_4$ is chosen such that it reflects the state
$\ket{L^\prime}$ towards the mirror $M_3$, leading to the mirror $M_5$ and finally onto the beam-splitter $BS_5$. From the 
beam-splitter $BS_4$, any state, other than $\ket{L^\prime}$, is transmitted to the detector $D_5$. Moving onto the beam-splitter
$BS^\prime_4$, it is so chosen that only a state $\ket{R}$ is reflected towards the mirror $M^\prime_4$, leading to the 
beam-splitter $BS_5$. A state, other than $\ket{R}$ is transmitted to the detector $D^\prime_5$. The inputs of the 
beam-splitter $BS_5$ are thus $\ket{R}$ and $\ket{L^\prime}$. To achieve the desired postselection, we must choose 
$BS_5$ so that it allows only the state $\ket{R}\ket{L^\prime}$ towards the beam-splitter $BS_6$.
Any other state is sent to the detector $D_4$. It is to be remembered that the beam-splitter $BS_6$ also receives the 
state $\ket{L}\ket{R^\prime}$. We choose the beam-splitter $BS_6$ in a way that only a state $\ket{L}\ket{R^\prime}-
\ket{R}\ket{L^\prime}$ is allowed towards the mirror $M_6$ and finally towards the beam-splitter $BS^\prime_7$.
A state, other than $\ket{L}\ket{R^\prime}-\ket{R}\ket{L^\prime}$ is sent towards the detector $D^\prime_3$ from $BS_6$.
Now let us go back to the states $\ket{L}\ket{L^\prime}$ and $\ket{R}\ket{R^\prime}$ proceeding towards the beam splitter
$BS_7$. This beam-splitter is such that the state $\ket{L}\ket{L^\prime}+\ket{R}\ket{R^\prime}$ is transmitted towards
the beam-splitter $BS^\prime_7$ and any other state is transmitted towards the detector $D_3$. Finally the beam-splitter
$BS^\prime_7$ sends the state $\ket{L}\ket{L^\prime}+\ket{R}\ket{R^\prime}+\ket{L}\ket{R^\prime}-\ket{R}\ket{L^\prime}$
towards the two-photon detector $D_1$ while any other state is sent towards the detector $D^\prime_1$.

Thus by selecting the clicks of the detector $D_1$ alone, we can successfully postselect the 
desired state. To appreciate the working of this arrangement, let us define two circular polarization bases
\begin{eqnarray}
 \ket{+}&=&\frac{1}{\sqrt{2}}(\ket{H}+ i \ket{V})\nonumber,\\
 \ket{-}&=&\frac{1}{\sqrt{2}}(\ket{H}- i \ket{V})
\end{eqnarray}
and
\begin{eqnarray}
 \ket{+^\prime}&=&\frac{1}{\sqrt{2}}(\ket{H^\prime}+ i \ket{V^\prime})\nonumber,\\
 \ket{-^\prime}&=&\frac{1}{\sqrt{2}}(\ket{H^\prime}- i \ket{V^\prime})
\end{eqnarray}
and two operators
\begin{eqnarray}
 \sigma_z&=&\ket{+}\bra{+}-\ket{-}\bra{-}\nonumber,\\
 \sigma_{z^\prime}&=&\ket{+^\prime}\bra{+^\prime}-\ket{-^\prime}\bra{-^\prime}.
\end{eqnarray}

To find out which arms the two photons passed through, we must perform measurements of the observables 
$\Pi_L=\ket{L}\bra{L}$, $\Pi_R=\ket{R}\bra{R}$, $\Pi_{L^\prime}=\ket{L^\prime}\bra{L^\prime}$ and
$\Pi_{R^\prime}=\ket{R^\prime}\bra{R^\prime}$ in the left and the right arms of the unprimed and the primed
halves of the arrangement, respectively. However, we 
must also ensure that the measurements do not disturb the initial state much, so that the probability statistics
for the desired postselected state is minimally altered. To this end, the above observables must be weakly 
measured by placing weak detectors in the four arms.

We must also ensure that the observed phenomenon is unambiguously quantum and cannot be explained
by classical theory which, as we have briefly discussed earlier has raised suspiscion in previous works~\cite{Atherton2015}.
In our setup, the weak detection of an operator, say $(\Pi_L)^w$, begins by weakly coupling the system, in 
our case, the photon, with a second system or a meter. In terms of physics, this is usually done through an interaction
Hamiltonian~\cite{AAV1988}, resulting in a unitary acting jointly on the system and the meter. The meter consists of a 
two level system, defined in 
$\{\ket{0}_m,\ket{1}_m\}$ basis, and initialized in the $\ket{0}_m$ state. Following the technique laid out in
Ref.~\cite{Kim2018}, we execute a joint unitary $U_{\Pi_L}$ acting on the unprimed photon and the meter as
\begin{eqnarray}
 U_{\Pi_L}= && (I-\Pi_L)\otimes \ket{H}\bra{H}\otimes I \otimes I \otimes I \nonumber\\ 
 &+&  I\otimes \ket{V}\bra{V} \otimes I \otimes I\otimes I  +\Pi_L \otimes \ket{H}\bra{H}\nonumber\\
 &\otimes& I \otimes I\otimes R^{-1}(\theta_g) Z R(\theta_g).\label{uni1}
\end{eqnarray}
Each term of the above unitary is a tensor product of operators, acting respectively on the path degree of freedom
of the unprimed photon, its polarization, path degree of freedom of the primed photon, its polarization and the 
meter. The operator $Z=\ket{0}_m\bra{0}_m-\ket{1}_m\bra{1}_m$. $R(\theta_g)$ is defined such that 
\begin{eqnarray}
R(\theta_g)\ket{0}_m&=&\cos{2\theta_g}\ket{0}_m+\sin{2\theta_g}\ket{1}_m,\nonumber\\ 
R(\theta_g)\ket{1}_m&=&\sin{2\theta_g}\ket{0}_m-\cos{2\theta_g}\ket{1}_m,
\end{eqnarray}
with $g=4\theta_g$. The coupling between the photon and the meter is made weak by tuning for low values of $g$ so 
that only the terms till the first order in $g$ make significant contribution. Under the postselection, given by
Equation~(\ref{pos}), the meter is thrown into a state
\begin{equation}
 \ket{\Phi}_m= \bra{\Psi_f}U_{\Pi_L}\ket{\Psi_i}\ket{0}_m.\label{fms1}
\end{equation}
Using Equations~(\ref{prs}),~(\ref{pos}) and~(\ref{uni1}), Equation~(\ref{fms1}) assumes the form
\begin{equation}
 \ket{\Phi}_m\propto \ket{0}_m+g (\Pi_L)^w \ket{1}_m,\label{fms2}
\end{equation}
with $(\Pi_L)^w=\frac{\bra{\Psi_f}\Pi_L\ket{\Psi_i}}{\bra{\Psi_f}\Psi_i\rangle}=1$ being the weak value of $\Pi_L$.
Equation~(\ref{fms2}) is reached by using the knowledge from theory of weak values that for successful postselections,
the meter 
state shows an average shift, proportional to the weak value of the observable being measured~\cite{AAV1988, Kim2018}.

Similarly a weak measurement of $\Pi_R$ can be carried out by choosing a global unitary, obtained by substituting
$\Pi_L$ by $\Pi_R$ in Equation~(\ref{uni1}). A similar reasoning can be followed for the primed projection operators.
In addition to~\cite{Kim2018}, a general framework to construct a joint unitary evolution and perform measurement
using weak detectors can be found in~\cite{Pryde2005}. The weak values obtained are 

\begin{eqnarray}
 (\Pi_L)^w=\frac{\bra{\Psi_f}\Pi_L\ket{\Psi_i}}{\bra{\Psi_f}\Psi_i\rangle}=1,\nonumber\\
 (\Pi_R)^w=\frac{\bra{\Psi_f}\Pi_R\ket{\Psi_i}}{\bra{\Psi_f}\Psi_i\rangle}=0,\nonumber\\
 (\Pi_{L^\prime})^w=\frac{\bra{\Psi_f}\Pi_{L^\prime}\ket{\Psi_i}}{\bra{\Psi_f}\Psi_i\rangle}=0,\nonumber\\
 (\Pi_{R^\prime})^w=\frac{\bra{\Psi_f}\Pi_{R^\prime}\ket{\Psi_i}}{\bra{\Psi_f}\Psi_i\rangle}=1.
 \label{photon_swap}
\end{eqnarray}
Thus the photon at the unprimed input port must have traveled through the left arm of the unprimed
half of the setup while the photon at the primed input port must have traveled through the right
arm of the primed half of the setup when the postselection of $\ket{\Psi_f}$ is done.

 \begin{figure}
  \includegraphics[scale=0.38]{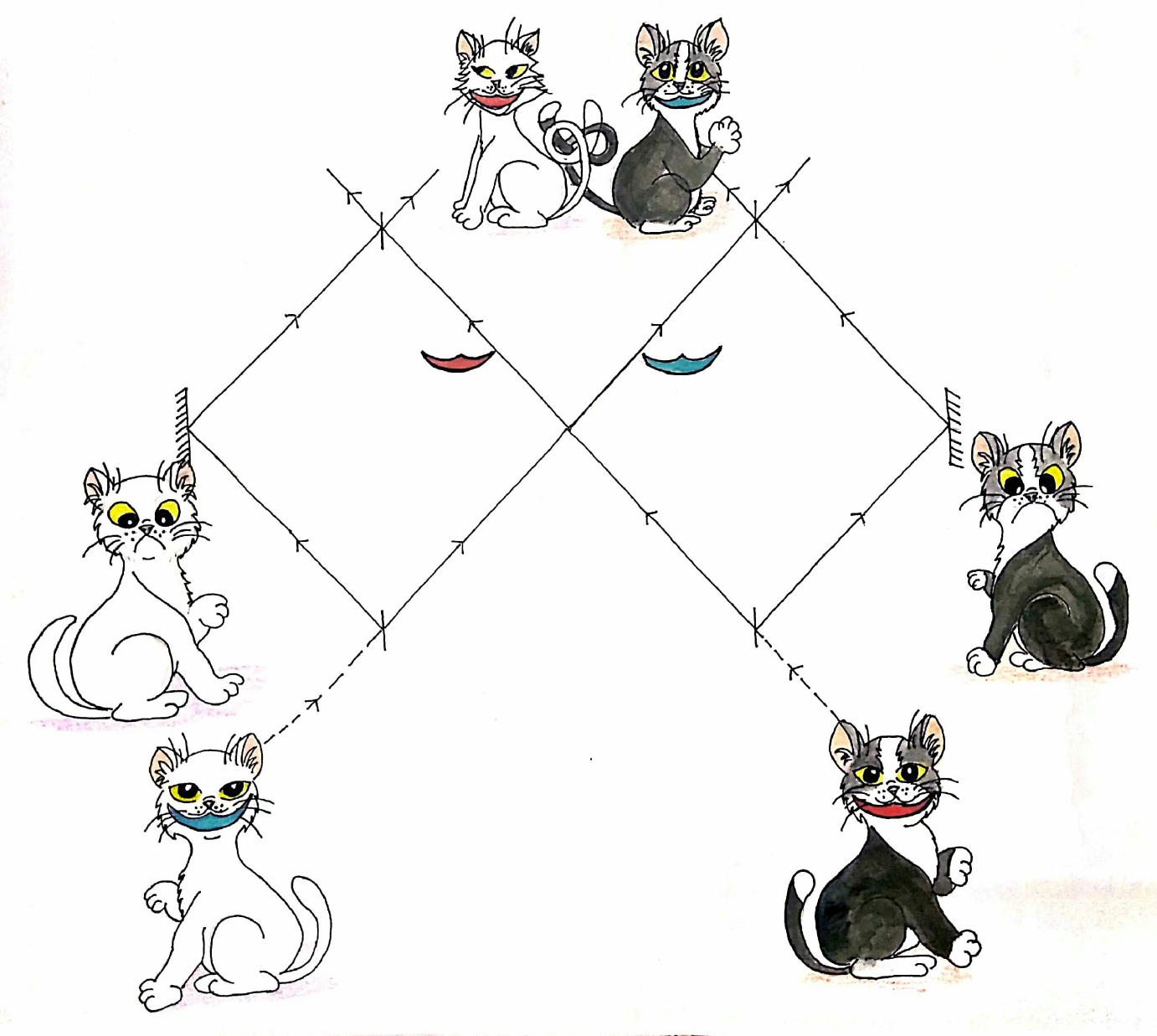}
  \caption{Two Quantum Cheshire Cats one white, with blue grin, and the other black, with a red grin, enter the arrangement. The 
  paths taken by the cats and their grins are shown. Each grin decouples from its respective cat and then recouples
  with the other cat. The final result is the exchange of the two grins and the formation of an entangled state, made up
  of a white cat, with red grin, and a black cat, with blue grin, both the cats being spatially separated from each other.}
  \label{cartoon}
 \end{figure}

Similarly, the polarizations can be detected in the left and the right arms
by respectively performing weak measurements of the operators $\sigma_z^L=\Pi_L\otimes\sigma_z$,
$\sigma_z^R=\Pi_R\otimes\sigma_z$,
$\sigma_{z^\prime}^{L^\prime}=\Pi_{L^\prime}\otimes\sigma_{z^\prime}$ and
$\sigma_{z^\prime}^{R^\prime}=\Pi_{R^\prime}\otimes\sigma_{z^\prime}$ to detect the path of the polarization of the  photons 
within the  designated arms of the
interferometer. To understand these operators, let us take the example of say $\sigma_z^R=\Pi_R\otimes\sigma_z$. On the 
unprimed side of the setup, the beam-splitter $BS_1$ sends the photon and the polarization it carries
into either $\ket{L}$ or $\ket{R}$. The polarization on the unprimed side is acted upon by unprimed operators. Thus to 
find the polarization on the right arm of the unprimed part, it is necessary to measure the operator 
$\sigma_z^R=\Pi_R\otimes\sigma_z$. A similar reasoning can be reached for the other choices of the operators. 
To obtain each weak value we once again include a meter and interact it with the relevant photon by means
of a global unitary. For example, the unitary for weakly measuring $\sigma_z^L$ should have the form   
\begin{eqnarray}
 U_{\sigma_z^L}= && \frac{1}{2}[(I-\Pi_L)\otimes \sigma_z\otimes I \otimes I \otimes I \nonumber\\ 
 &+&  I\otimes \sigma_z \otimes I \otimes I\otimes I +\Pi_L \otimes \sigma_z\nonumber\\
 &\otimes& I \otimes I\otimes R^{-1}(\theta_g) Z R(\theta_g)].\label{uni2}
\end{eqnarray}
Similarly other unitaries for the weak measurements of the positions of circular polarization for the unprimed and 
primed photons can be defined.

These weak values turn out to be
\begin{eqnarray}
 (\sigma_z^L)^w=\frac{\bra{\Psi_f}\Pi_L\otimes\sigma_z\ket{\Psi_i}}{\bra{\Psi_f}\Psi_i\rangle}=0,\nonumber\\
 (\sigma_z^R)^w=\frac{\bra{\Psi_f}\Pi_R\otimes\sigma_z\ket{\Psi_i}}{\bra{\Psi_f}\Psi_i\rangle}=1,\nonumber\\
 (\sigma_{z^\prime}^{L^\prime})^w=\frac{\bra{\Psi_f}\Pi_{L^\prime}\otimes\sigma_{z^\prime}\ket{\Psi_i}}{\bra{\Psi_f}\Psi_i\rangle}=1,
 \nonumber\\
 (\sigma_{z^\prime}^{R^\prime})^w=\frac{\bra{\Psi_f}\Pi_{R^\prime}\otimes\sigma_{z^\prime}\ket{\Psi_i}}{\bra{\Psi_f}\Psi_i\rangle}=0,
 \label{pol_swap}
\end{eqnarray}
which means that the circular polarization of the photon at the unprimed input port must 
have traveled via the right arm of the unprimed half of the arrangement and the circular
polarization of the photon at the primed input port must have journeyed via the left arm of 
the primed half of the setup for all final outcomes $\ket{\Psi_f}$. Equations~(\ref{photon_swap}) 
and~(\ref{pol_swap}) jointly demonstrate that, under the above postselection, the unprimed photon
at some stage was at the beam-splitter $BS_2$  but its circular polarization was at the beam-splitter $BS^\prime_2$.
Similarly, the primed photon reached the beam-splitter $BS^\prime_2$ while its circular 
polarization went to the beamsplitter $BS_2$, for the postselected state $\ket{\Psi_f}$. In other words, recombination
of the photons and their polarizations do occur, but in the case of the above mentioned postselected states,
the unprimed photon recombined with the primed polarization while the primed photon recombined with the unprimed
polarization. Thus we have exchanged the grins of two Quantum Cheshire Cats.

We would also emphasize on the fact that the above conclusions are based on defining quantum observables
and modelling each of their measurement involving a system-meter interaction. Although classical treatment of the 
phenomenon may yeild strikingly similar results, it would not clarify with precise definitions what quantum variables
would be measured and how. This would suggest that the phenomenon we describe is essentially quantum in nature.

It is to be noted that to execute the process of postselection, culminating in the detection of two photons at 
$D_1$, the two photons come together once again. As a result we cannot comment on the whether the circular 
polarizations are once again exchanged and whether they return to their original photons. But it is equally
true that for the successful postselection of the state $\ket{\Psi_f}$, each circular polarization at 
certain stage of the photons' progress through the interferometer, say at $BS_2$ and $BS^\prime_2$, associate 
itself with the other photon.

Before we conclude, we would like to comment that the ambiguity of the association of photons
with their properties is closely related to the indistinguishability of quantum particles such as photons. It
is therefore especially interesting that the present idea does not rely on actual photon bunching effects. However,
some  important new insights about photon indistinguishability could be gained from the present results by comparing 
them a bit more closely to related results observed when photon bunching effects are present.
In  Ref. \cite{Hofmann2002} there has been a proposal  for the realization of an entanglement filter and
this has been realized 
experimentally in Ref. \cite{Okamoto2009}. In that scenario, two photons are also kept on separate outer paths while the
inner paths do cross. It should be noted that the entanglement filter is different from our scenario,
because the intended effect in those papers is the elimination of
single photon occupation in the inner paths by additional photon bunching. However, in our double Cheshire Cat
setup we achieve the
elimination of two photons arriving at the same side of the interferometer by suitable post-selection. 
It seems to be equally important for both cases that the
indistinguishability of photons makes it difficult to identify photons by keeping track of their spatial modes, as
observed in the instances of two photons leaving the optical circuit on the same side. To describe photons
as distinct physical objects, we therefore need to make an effort to keep photons in separate modes. The exchange
of polarization in the present scenario seems to demonstrate that non-trivial multi-photon quantum interferences can
occur without any actual photon bunching. This is probably mediated by the post-selected state. In future, it may be
worth exploring how the present exchange of photon polarization relates to the well-established realizations of
photon-photon interactions mediated by quantum interferences between the indistinguishable cases of photon exchange
and no photon exchange.


{\it Conclusions.--}
To summarize, we have proposed a thought experiment in which the circular polarizations of two photons can
be exchanged using interferometric arrangements. 
The setup for executing this process is based on the original
Quantum Cheshire Cat setup where the circular polarization can be separated from the photon
for suitable postselected states. Our method strives to decouple the polarization and the photon, and 
replace the original polarization with another that was previously associated with a 
different photon. This polarization in turn associates itself with the second photon. 
 Our Cheshire Cat experiment not only separates a particle from its properties but also
 causes one particle to acquire a property previously associated with another particle.

The implications for the exchanging of photon polarization are significant.  Our thought-experiment allows us to think about the deep insight on the relation between quantum measurement itself and revealing reality of quantum state. Firstly, it challenges the 
notion that a property must faithfully `belong' to a particular physical system. In the realm of the 
quantum systems, this `belongingness' is certainly very capricious with properties belonging to independent
physical systems getting interchanged. This raises the question of reality of observables as a much deeper level. 
In our case the photon is decoupled from its circular 
polarization and then replaced with the circular polarization from a second photon. The second point to note
is  that entanglement plays a crucial 
role in the realization of this exchange process. As discussed before, the exchange is successful only
when a certain outcome is attained at the end. This so happens that this outcome is an entangled state. It 
is possible that with different pre and postselected states one can also observe the exchange of grins.
In future, it may be worth exploring if one can also exchange grins of two Cheshire cats with a postselected
state that is not entangled. This may  indeed mean that one could swap the polarizations of two photon without
ever interacting. These may provide new insights in foundations of quantum mechanics, quantum information as
well as in technological applications.

\section{Acknowledgements}

The authors thank Sreeparna Das for her help with Fig.~(\ref{cartoon}). This work has been funded by 
J C Bose Fellowship from Department of Science and Technology (DST), India, under Grant No. JCB/2018/000038
(2019-2024).

\newpage

\end{document}